\documentclass[12pt]{article}
\PassOptionsToPackage{natbib=true}{biblatex}
\usepackage[latin9]{inputenc}
\usepackage{amsmath}
\usepackage{amssymb}
\usepackage{esint}
\usepackage[pdfusetitle,
 bookmarks=true,bookmarksnumbered=false,bookmarksopen=false,
 breaklinks=false,pdfborder={0 0 0},pdfborderstyle={},backref=false,colorlinks=false]
 {hyperref}

\makeatletter
\numberwithin{equation}{section}

\usepackage{amsfonts}
\AddToHook{package/after/biblatex}{%
  \AtEveryBibitem{%
    \ifentrytype{article}
      {%
        \clearfield{title}%
        \clearfield{subtitle}%
        \clearfield{titleaddon}%
      }
      {}%
  }%
  \renewbibmacro{in:}{%
    \ifentrytype{article}
      {}
      {\printtext{\bibstring{in}\intitlepunct}}%
  }%
}

\makeatother

\usepackage[style=numeric-comp,maxbibnames=5, firstinits, sorting=none]{biblatex}
\begin{document}
\begin{flushright}
FIAN/TD/2026-19
\par\end{flushright}

\vspace{1cm}
 
\begin{center}
{\large\textbf{Unfolded Schwinger\textendash Dyson Hierarchy for Vertex
Functions}}{\large\par}
\par\end{center}

\begin{center}
\vspace{0.6cm}
 \textbf{Nikita Misuna}\\
 \vspace{0.5cm}
 \textit{Tamm Department of Theoretical Physics, Lebedev Physical
Institute,}\\
 \textit{Leninsky prospekt 53, 119991, Moscow, Russia}\\
 
\par\end{center}

\begin{center}
\vspace{0.4cm}
misuna@lpi.ru \\
 \vspace{0.6cm}
\par\end{center}
\begin{abstract}
We propose a formulation of a quantum field theory within the framework
of the unfolded dynamics approach, originally developed in higher-spin
gauge theory. We construct an unfolded system that describes a hierarchy
of Schwinger\textendash Dyson equations for vertex functions of a
scalar QFT in terms of unfolded first-order equations. A distinctive
feature of the constructed system is the appearance of two additional
evolution parameters responsible for the WKB and RG behaviors. As
an illustration, a one-loop analysis of the theory with quartic self-interaction
is carried out.

\newpage{}

\tableofcontents{}
\end{abstract}

\section{Introduction}

Quantum field theory (QFT) remains the main tool for constructing
and investigating models of fundamental interactions. Despite the
impressive progress in both technical and conceptual aspects of QFT
made over more than half a century, many problems remain open such
as divergences, quantum triviality, non-perturbative foundations,
etc. This calls for a search for new approaches to QFT. Taking into
account the ever-increasing role of symmetry considerations, it is
natural to expect that future advances will be closely related to
the development of our ability to deal with symmetries.

A clear illustration of the importance of symmetries in modern theoretical
physics is provided by higher-spin gravity theories (for recent progress
in the field, see \citep{Kirakosiants:2025gpd,Didenko:2026nag,Giombi:2026sqa,Kirakosiants:2026spd,Didenko:2026apw,De:2026shn,Misuna:2026has,Serrani:2026azw,Didenko:2026nip,Zaigraev:2026jvu,Cho:2026mjp,Skvortsov:2026ofl,Faliakhov:2026yum,Zaigraev:2026xzr,Kozaki:2026rdc,Boulanger:2026wle,Karapetyan:2026cbh,Batyaev:2026aaw,Lang:2026inn,Valerev:2026wos}).
These theories are to a large extent determined by infinite-dimensional
higher-spin algebras \citep{Fradkin:1986ka,Vasiliev:1986qx,Vasiliev:2004cm,Eastwood:2002su,Boulanger:2013zza,Joung:2014qya,Campoleoni:2021blr,Ponomarev:2026azo}.
The source of these infinite-dimensional symmetries is the infinite
higher-spin spectrum that consists of massless fields with unrestricted
spins. To efficiently process such systems, the\emph{ unfolded dynamics
approach} \citep{Vasiliev:2005zu} has been developed \citep{Vasiliev:1988xc},
which allowed for a formulation of a closed generating system for
a nonlinear higher-spin theory \citep{Vasiliev:1990en,Vasiliev:1992av}.
The unfolded dynamics approach reformulates the content of a classical
theory in the form of first-order differential equations on exterior
forms that represent unfolded fields. The unfolded formulation enjoys
manifest diffeomorphism and gauge invariance and provides an efficient
organization of the theory's degrees of freedom: an infinite (in dynamical
field theories) number of unfolded fields parameterizes all d.o.f.
and determines their transformation properties under the theory's
symmetries. Moreover, background universality of the unfolded dynamics
approach allows one to set the same unfolded system on different background
manifolds, the property that underlies superspace extensions of supersymmetric
theories \citep{Ponomarev:2010ji,Misuna:2013ww,Misuna:2026bhy} and
holographic dualities \citep{Vasiliev:2012vf}. A quantization scheme
for classical unfolded theories was put forward in \citep{Misuna:2022cma}
(see also \citep{Lyakhovich:2004xd,Barnich:2004cr,Kazinski:2005eb,Lyakhovich:2005mk,Lyakhovich:2006sc,Kaparulin:2010ywf,Kaparulin:2011xy,Kaparulin:2011aa}
for related discussions of quantization of non-Lagrangian field theories).

Beyond higher-spin gravity models, not so many examples of unfolded
nonlinear lower-spin field theories are available \citep{Misuna:2022cma,Misuna:2024ccj,Misuna:2024dlx}
and even those look cumbersome and excessive due to the finiteness
of their field spectra. This suggests that a promising application
of the unfolded dynamics approach should be related to models with
infinite spectra.

A relatively simple yet rich and substantive example of such models
is provided by scalar QFT viewed as a theory of an infinite tower
of vertex functions governed by the hierarchy of Schwinger\textendash Dyson
equations. The aim of the present paper is to construct and analyze
an unfolded system for this model. Unfolded Schwinger\textendash Dyson
equations have already been considered in \citep{Misuna:2022cma}.

In that work, however, vertex functions were treated in the conventional
way, as spacetime multipoint functions related by differential Schwinger\textendash Dyson
equations. The resulting unfolded system was therefore traditional
in the sense that it expressed spacetime derivatives in terms of unfolded
mean fields. Here, we adopt a completely different point of view and
treat vertex functions as symmetric tensors on an infinite-dimensional
configuration space whose coordinates are the fields. Accordingly,
the unfolded equations for vertex functions become variational in
nature and are supplemented by unfolded equations for two additional
scalar parameters governing WKB and RG flows.

The paper is organized as follows. In Section \ref{SECTION_UNFOLDING},
we give a brief review of the unfolded dynamics approach. In Section
\ref{SECTION_QUANT_EFF_ACT}, we recall the necessary facts about
the quantum effective action, Schwinger\textendash Dyson equations
and functional RG. Then, in Section \ref{SECTION_UNFOLD_SD}, we construct
an unfolded system for the vertex functions of scalar QFT. In Section
\ref{SECTION_ONE_LOOP} we give an example of the one-loop analysis
of the $\lambda\phi^{4}$ theory. Finally, we present our conclusions
in Section \ref{SECTION_CONCLUSION}.

\section{Basics of unfolding\label{SECTION_UNFOLDING}}

Within the unfolded dynamics approach \citep{Vasiliev:1988xc,Vasiliev:2005zu}
(see also \citep{Sullivan:1977pdi,DAuria:1982uck,Castellani:2025axf}
for related concepts), a classical field theory is formulated in the
form of unfolded equations
\begin{equation}
\mathrm{d}W^{A}(z)+G^{A}(W)=0,\label{unf_eq}
\end{equation}
representing first-order equations on unfolded fields $W^{A}(z)$
that are exterior forms on some background manifold ${\cal M}^{D}$
with local coordinates $z^{i=\overline{1,D}}$ and an exterior derivative
$\mathrm{d}=\mathrm{d}z^{i}\frac{\partial}{\partial z^{i}}$. The
collective multi-index $A$ denotes all indices carried by the unfolded
field, and $G^{A}(W)$ is constructed from the exterior products of
the unfolded fields. For each unfolded field $W^{A}$, there is an
associated unfolded equation \eqref{unf_eq}.

The identity $\mathrm{d}^{2}\equiv0$ requires $G^{A}$ to be closed,
$\mathrm{d}G^{A}=0$, which by means of \eqref{unf_eq} is reformulated
as a consistency constraint
\begin{equation}
G^{B}\wedge\dfrac{\partial G^{A}}{\partial W^{B}}=0.\label{unf_consist}
\end{equation}

A consistent system of unfolded equations \eqref{unf_eq} is manifestly
invariant under infinitesimal unfolded gauge transformations
\begin{equation}
\delta W^{A}=\mathrm{d}\varepsilon^{A}(z)-\varepsilon^{B}\wedge\dfrac{\partial G^{A}}{\partial W^{B}}.\label{unf_gauge_transf}
\end{equation}
According to this, every $(n>0)$-form field $W^{A}$ gives rise to
an unfolded gauge symmetry with a parameter $\varepsilon^{A}(z)$
being a $(n-1)$-form, while 0-form fields are transformed only under
the gauge symmetries of the 1-form fields due to the second term in
\eqref{unf_gauge_transf}.

Unfolded formulations of conventional field theories involve only
1-forms associated with gauge d.o.f. and 0-forms associated with physical
d.o.f. (because, in the linear approximation, they are transformed
only under vacuum global symmetries).\footnote{Unfolded higher-form fields, especially relevant to the action principle
problem, are discussed in, e.g., \citep{Vasiliev:2005zu,Vasiliev:2015mka,Didenko:2015pjo,Didenko:2017qik,Tarusov:2021fdk}.} Thus, unfolding of a dynamical field theory requires introduction
of an infinite number of unfolded 0-form fields. Then an unfolded
system \eqref{unf_eq} describes some theory of dynamical (primary)
fields in terms of infinite towers of all their differential descendants.
These towers are by construction manifestly covariant w.r.t. all unfolded
symmetries \eqref{unf_gauge_transf}. Furthermore, because the unfolded
equations are formulated in terms of exterior forms, an unfolded system
also possesses a manifest diffeomorphism invariance on ${\cal M}^{D}$.
Even more, because \eqref{unf_consist} is insensitive to the choice
of ${\cal M}^{D}$ (provided one does not use vanishing of ($n>D$)-forms
when solving for the consistency), a consistent unfolded system can
be placed on an arbitrary background manifold. This way one unfolded
system generates different theories in different spaces, but all having
the same spectra of symmetries and physical d.o.f. (since these are
described by 0-forms which do not depend on the background manifold).
This feature of background universality allows one to study systematically
superspace extensions of supersymmetric theories \citep{Ponomarev:2010ji,Misuna:2013ww,Misuna:2026bhy}
and holographic dualities \citep{Vasiliev:2012vf}.

Let us illustrate the structure of an unfolded system using as an
example a free scalar field in $D$-dimensional Minkowski space \citep{Shaynkman:2000ts}.
To describe the background, one first introduces a 1-form
\begin{equation}
\Omega=e^{a}P_{a}+\omega^{a,b}M_{a,b}
\end{equation}
that takes values in the Poincaré algebra (here $P_{a}$ and $M_{a,b}$
are generators of translations and rotations, while $e^{a}$ and $\omega^{a,b}$
are vielbein and Lorentz connection fields; $a,b=0,...,D-1$ are indices
of the fiber over a background manifold) and imposes the Maurer\textendash Cartan
equation on it
\begin{equation}
\mathrm{d}\Omega+\frac{1}{2}[\Omega\wedge\Omega]=0.\label{Maurer-Cartan_eq}
\end{equation}
A scalar field in the unfolded dynamics approach corresponds to an
infinite set of unfolded fields that are symmetric tensors $\Phi_{a_{1}...a_{n}}$
of all ranks subject to unfolded equations
\begin{equation}
\mathrm{d}\Phi_{a_{1}...a_{n}}+\omega_{(a_{1}}{}^{b}\Phi_{a_{2}...a_{n})b}-e^{b}\Phi_{ba_{1}...a_{n}}=0.\label{scalar_eq}
\end{equation}

An unfolded system \eqref{Maurer-Cartan_eq}, \eqref{scalar_eq} is
consistent. Corresponding unfolded gauge symmetry \eqref{unf_gauge_transf}
has the form
\begin{align}
\delta e^{a} & =\mathrm{d}\varepsilon^{a}+\omega^{a}{}_{b}\varepsilon^{b}-\varepsilon^{a}{}_{b}e^{b},\label{gauge_1}\\
\delta\omega^{a,b} & =\mathrm{d}\varepsilon^{a,b}+\omega^{a}{}_{c}\varepsilon^{c,b},\label{gauge_2}\\
\delta\Phi_{a_{1}...a_{n}} & =-\varepsilon_{(a_{1}}{}^{b}\Phi_{a_{2}...a_{n})b}+\varepsilon^{b}\Phi_{ba_{1}...a_{n}}\label{gauge_3}
\end{align}
and describes local Poincaré transformations in the fiber space.

At this stage, the background manifold remains unfixed and for any
choice there is still an infinite-dimensional diffeomorphism symmetry.
But in order to see that \eqref{Maurer-Cartan_eq}, \eqref{scalar_eq}
indeed describes a scalar field, one has to fix the background and
coordinates on it which then allows one to analyze the system. Thus,
we fix ${\cal M}^{D}=\mathbb{R}^{1,D-1}$ and choose the simplest
non-degenerate solution to \eqref{Maurer-Cartan_eq}, which is global
Cartesian coordinates
\begin{equation}
e_{0}^{a}=\mathrm{d}x^{a},\quad\omega_{0}^{a,b}=0.\label{Cartesian}
\end{equation}
The choice of the background solution breaks an infinite-dimensional
diffeomorphism symmetry down to the global Poincaré symmetry generated
by those gauge transformations \eqref{gauge_1}-\eqref{gauge_2} that
leave \eqref{Cartesian} intact:
\begin{align}
 & \delta\omega_{0}^{a,b}=0,\quad\Rightarrow\quad\varepsilon_{0}^{a,b}=\xi^{a,b},\label{global_1}\\
 & \delta e_{0}^{a}=0,\quad\Rightarrow\quad\varepsilon_{0}^{a}=\xi^{a}+\xi^{a}{}_{b}x^{b},\label{global_2}
\end{align}
with constant global symmetry parameters $\xi^{a}$, $\xi^{a,b}$.

In Cartesian coordinates, \eqref{scalar_eq} boils down to 
\begin{equation}
\frac{\partial}{\partial x^{b}}\Phi_{a_{1}...a_{n}}=\Phi_{ba_{1}...a_{n}},
\end{equation}
so one has
\begin{equation}
\Phi_{a_{1}...a_{n}}=\frac{\partial}{\partial x^{a_{1}}}...\frac{\partial}{\partial x^{a_{n}}}\phi(x).\label{scalar_descendants}
\end{equation}
Thus, an infinite tower of unfolded fields indeed encodes all differential
descendants of a primary scalar field $\phi(x)$. The transformations
\eqref{gauge_3} with parameters \eqref{global_1}, \eqref{global_2}
then define a representation of the global Poincaré algebra on a module
spanned by $\Phi_{a_{1}...a_{n}}$.

The system \eqref{Maurer-Cartan_eq}, \eqref{scalar_eq} does not
impose any constraints on the primary scalar $\phi(x)$. Due to this,
it is referred to as an off-shell unfolded system. But if one now
declares that $\Phi_{a_{1}...a_{n}}$ are traceless $\eta^{bc}\Phi_{bca_{1}...a_{n-2}}=0$
(note that this does not spoil the consistency, which is crucial),
then \eqref{scalar_eq} via \eqref{scalar_descendants} entails the
massless Klein\textendash Gordon equation
\begin{equation}
\square\phi(x)=0.
\end{equation}
In this case, the unfolded system is referred to as an on-shell one.

\section{Quantum effective action and vertex functions\label{SECTION_QUANT_EFF_ACT}}

Here we briefly recall the necessary information about a one-particle
irreducible quantum effective action which will be re-analyzed within
the unfolded framework in the next Section.

One starts with a generating functional of a Euclidean scalar QFT
(a single real scalar field) of the form
\begin{equation}
Z[J]=\int{\cal D}\varphi\exp\left\{ -\frac{1}{\hbar}S[\varphi]+\frac{1}{\hbar}\int\mathrm{d}^{D}x\varphi(x)J(x)\right\} 
\end{equation}
and introduces a generating functional of connected correlators as
\begin{equation}
W[J]=\hbar\ln Z[J].
\end{equation}
The 1PI scalar effective action $\Gamma[\phi]$ is the Legendre transform
of $W[J]$
\begin{equation}
\Gamma[\phi]=\int\mathrm{d}^{D}x\phi(x)J(x)-W[J],
\end{equation}
so that for the classical field $\phi(x)=\bigl\langle\hat{\varphi}(x)\bigr\rangle$
and the source $J(x)$ one has
\begin{equation}
\phi=\frac{\delta W}{\delta J},\quad J=\frac{\delta\Gamma}{\delta\phi}.
\end{equation}
Then $Z$ can be expressed in terms of the classical field as
\begin{equation}
Z[J(\phi)]=\exp\{\frac{1}{\hbar}W[J(\phi)]\}=\exp\{\frac{1}{\hbar}\int\mathrm{d}^{D}x\phi(x)J(x)-\frac{1}{\hbar}\Gamma[\phi]\}=\exp\{\frac{1}{\hbar}(\int\mathrm{d}^{D}x\phi(x)\frac{\delta}{\delta\phi(x)}-1)\Gamma[\phi]\}.\label{Z_of_phi}
\end{equation}
In addition,
\begin{equation}
\frac{\delta}{\delta J(x)}=\int\mathrm{d}^{D}y\frac{\delta\phi(y)}{\delta J(x)}\frac{\delta}{\delta\phi(y)}=\int\mathrm{d}^{D}y(\frac{\delta J(x)}{\delta\phi(y)})^{-1}\frac{\delta}{\delta\phi(y)}=\int\mathrm{d}^{D}y(\frac{\delta^{2}\Gamma}{\delta\phi(x)\delta\phi(y)})^{-1}\frac{\delta}{\delta\phi(y)}.\label{J_derivative}
\end{equation}

From the vanishing of the functional integral of a total derivative
\begin{equation}
\int{\cal D}\varphi\frac{\delta}{\delta\varphi(x)}\exp\left\{ -\frac{1}{\hbar}S[\varphi]+\frac{1}{\hbar}\int\mathrm{d}^{D}y\varphi(y)J(y)\right\} =0,
\end{equation}
a functional Schwinger\textendash Dyson equation for correlation functions
follows
\begin{equation}
\frac{\delta S}{\delta\varphi(x)}[\hbar\frac{\delta}{\delta J}]Z[J]=J(x)Z[J].\label{SD_equation}
\end{equation}
This differential equation on $Z[J]$ can be considered as the definition
of a quantum field theory corresponding to the classical action $S[\varphi]$.
The functional Schwinger\textendash Dyson equation \eqref{SD_equation}
can be reformulated in terms of a quantum e.o.m. for a classical field
$\phi(x)$. To this end, one substitutes \eqref{J_derivative} and
\eqref{Z_of_phi} into \eqref{SD_equation}, arriving at
\begin{equation}
\frac{\delta S}{\delta\varphi(x)}[\phi(x)+\hbar\int\mathrm{d}^{D}yG(x,y)\frac{\delta}{\delta\phi(y)}]1=\frac{\delta\Gamma}{\delta\phi(x)},\label{rel_1}
\end{equation}
where $\frac{\delta}{\delta\phi(y)}$ acts to the right and the full
propagator is defined via
\begin{equation}
\int\mathrm{d}^{D}y\frac{\delta^{2}\Gamma}{\delta\phi(x)\delta\phi(y)}G(y,z)=\delta^{(D)}(x-z).\label{rel_2}
\end{equation}
An $n$-point vertex function represents an $n$-th variational derivative
of $\Gamma$ evaluated at the vacuum value of $\phi(x)$
\begin{equation}
\Gamma^{(n)}(x_{1},x_{2},...x_{n})=\frac{\delta^{n}\Gamma}{\delta\phi(x_{1})\delta\phi(x_{2})...\delta\phi(x_{n})}|_{\phi=\phi_{vac}}
\end{equation}
and determines an $n$-point amplitude contribution to the $S$-matrix.
The vacuum value $\phi_{vac}$ is determined by the quantum e.o.m.
\begin{equation}
\frac{\delta\Gamma}{\delta\phi(x)}[\phi_{vac}]=0,\label{quant_eom}
\end{equation}
which typically yields $\phi_{vac}=0$ if there is no spontaneous
symmetry breaking.

The crucial observation for our subsequent analysis is that the two
relations \eqref{rel_1}, \eqref{rel_2} can be seen as the \emph{definition}
of a QFT based on the classical action $S$. Taking variational derivatives
of \eqref{rel_1} at $\phi=\phi_{vac}$, one recovers all vertex functions
(except for the vacuum contribution $\Gamma[\phi_{vac}]$). Because
the system \eqref{rel_1}, \eqref{rel_2} is highly self-entangled,
one usually resorts to the WKB expansion.

This scheme can be easily generalized in order to incorporate scale
dependence of an effective action. In the functional RG (FRG) approach,
one introduces a regulator term
\begin{equation}
\Delta S_{R}=\frac{1}{2}\int\mathrm{d}^{D}x\mathrm{d}^{D}y\varphi(x)R_{\tau}(x,y)\varphi(y)
\end{equation}
into the classical action, where $R_{\tau}(x,y)$ is a $\tau$-scale-dependent
regulator function. In the FRG context, $\Delta S_{R}$ serves as
an infrared cutoff that freezes the low-energy fluctuations, but here
we do not specify its properties. In particular, below we use it as
a UV regulator rendering loop integrals finite. The appearance of
$\Delta S_{R}$ requires some modifications to the scheme we set forth.
Concretely, one now defines the scale-dependent effective action as
\begin{equation}
\Gamma_{\tau}[\phi]=\int\mathrm{d}^{D}x\phi(x)J(x)-W[J]-\Delta S_{R}.
\end{equation}
The Schwinger\textendash Dyson equation keeps its form \eqref{rel_1},
but now the propagator is defined as
\begin{equation}
\int\mathrm{d}^{D}y\{\frac{\delta^{2}\Gamma_{\tau}}{\delta\phi(x)\delta\phi(y)}+R_{\tau}(x,y)\}G_{\tau}(y,z)=\delta^{(D)}(x-z)\label{rel_2R}
\end{equation}
instead of \eqref{rel_2}. The scale dependence of $\Gamma_{\tau}$
is then determined by the Wetterich equation \citep{Wetterich:1992yh}
\begin{equation}
\frac{\partial}{\partial\tau}\Gamma_{\tau}=\frac{\hbar}{2}\int\mathrm{d}^{D}x\int\mathrm{d}^{D}yG_{\tau}(x,y)\frac{\partial}{\partial\tau}R_{\tau}(y,x).\label{Wetterich_eq}
\end{equation}
Taking variational derivatives of \eqref{Wetterich_eq}, one finds
$\tau$-flows of the vertex functions.

\section{Unfolding the Schwinger\textendash Dyson hierarchy\label{SECTION_UNFOLD_SD}}

Our goal is to formulate an unfolded system that encodes in an unfolded
fashion all vertex functions of a scalar QFT. As said in the Introduction,
we are not going to involve the unfolding of a classical field theory
and quantize it as was proposed in \citep{Misuna:2022cma}. Instead,
we aim at constructing a system that directly operates with vertex
functions.

To begin with, we need to define the counterparts of unfolded fields.
It will be more convenient to work in  momentum space, so we consider
vertex functions as symmetric tensors $\Gamma_{p_{1}...p_{n}}$ on
an infinite-dimensional manifold $\mathcal{C}$ (field configuration
space) where the fields $\phi^{p}=\phi(p)$ play the role of (an infinite
number of) coordinates. Thus, the system we are going to formulate
formally represents an unfolded field theory on an infinite-dimensional
background manifold.

Here and below we make use of DeWitt condensed notations, so that
the contraction of upper and lower indices denotes integration over
the corresponding momentum
\begin{equation}
A_{p}B^{p}=\int\frac{d^{D}p}{(2\pi)^{D}}A(p)B(p).
\end{equation}
Indices denoted with the same letter are always symmetrized
\begin{equation}
T_{p_{1}p_{2}...p_{n}}=T_{(p_{1}p_{2}...p_{n})}.
\end{equation}

We naturally introduce an infinite tower of unfolded vertex functions
(defined at an arbitrary value of $\phi$, not necessarily the vacuum
one)
\begin{equation}
\delta_{p_{2}}\Gamma_{p_{1}}=\Gamma_{p_{1}p_{2}},\quad\delta_{p_{3}}\Gamma_{p_{1}p_{2}}=\Gamma_{p_{1}p_{2}p_{3}},\quad...\quad\delta_{q}\Gamma_{p_{1}p_{2}...p_{n}}=\Gamma_{qp_{1}p_{2}...p_{n}},\label{var_Gamma}
\end{equation}
with
\begin{equation}
\delta_{p}\equiv\frac{\delta}{\delta\phi^{p}},\quad\delta_{p}\phi^{q}=(2\pi)^{D}\delta^{(D)}(p-q),
\end{equation}
which is obviously analogous to the infinite tower of unfolded descendants
\eqref{scalar_descendants} of the unfolded scalar field module. The
difference is that here a primary field is $\Gamma_{p}$, not an effective
action $\Gamma$ by itself. However, symmetry of $\Gamma_{p_{1}p_{2}}$
implies that $\Gamma_{p}$ is a total derivative (Helmholtz condition)
and thus implicitly introduces the effective action. Classical vertex
functions are obviously defined as
\begin{equation}
\delta_{q}S_{p_{1}...p_{n}}=S_{qp_{1}...p_{n}}.
\end{equation}

We need to implement the two key relations \eqref{rel_1}, \eqref{rel_2}
that determine a QFT. To this end, we interpret the transition from
the classical to the quantum e.o.m. performed by \eqref{rel_1} as
the result of an evolution generated by some operator $\hat{{\cal U}}_{\ell}$
with respect to a parameter $\ell$, which we refer to as \emph{WKB-time}.
Thus,
\begin{equation}
\Gamma_{p}(\ell)=S_{p}[\phi_{q}+\hbar\ell G_{qk}\frac{\delta}{\delta\phi_{k}}]1=\hat{{\cal U}}_{\ell}S_{p},\quad\hat{{\cal U}}_{\ell=0}=1.\label{evolution_def}
\end{equation}
In practice, $\ell$ is a dimensionless analog of Planck's constant:
it always appears multiplied by $\hbar$ in all expressions. However,
their formal roles are different. Looking ahead, $\hbar$ governs
the standard semiclassical expansion, whereas the $\ell$-evolution
must be solved for exactly at each order in $\hbar$ and does not
admit an expansion in $\ell$.

Another point that needs to be addressed is that the notion of evolution
in this context requires further clarification\footnote{The author is grateful to A.V. Korybut for an illuminating discussion
of this point.}. Usually, an evolution operator is required to obey a natural composition
law,
\begin{equation}
\hat{{\cal U}}(\ell_{3},\ell_{1})=\hat{{\cal U}}(\ell_{3},\ell_{2})\hat{{\cal U}}(\ell_{2},\ell_{1}).\label{evolution_composition}
\end{equation}
However, $\hat{{\cal U}}_{\ell}$ in \eqref{evolution_def} is defined
as the operator that evolves the system from a distinguished point
$\ell$=0 (a classical theory) to some $\ell\neq0$ (a quantized theory
with effective Planck's constant equal to $\hbar\ell$). Thus, in
order to satisfy the composition law \eqref{evolution_composition},
the definition \eqref{evolution_def} should be extended as follows
\begin{equation}
\hat{{\cal U}}(\ell+\Delta_{\ell},\ell)=\begin{cases}
\hat{{\cal U}}_{\Delta_{\ell}}, & \ell=0,\\
\exp\{\Delta_{\ell}\frac{\partial}{\partial\ell}\}, & \ell\neq0.
\end{cases}
\end{equation}

Now we need to formulate a first-order equation for $\ell$-evolution
of $\Gamma_{p}$ that has to be of the form
\begin{equation}
\frac{\partial}{\partial\ell}\Gamma_{p}=\hat{H}(\ell)\Gamma_{p},\quad\Gamma_{p}(\ell=0)=S_{p}\label{Schroed_Gamma}
\end{equation}
with some formal Hamiltonian $\hat{H}$. Given $\hat{H}$, the solution
to \eqref{Schroed_Gamma} is
\begin{equation}
\Gamma_{p}(\ell)=\hat{{\cal T}}_{\ell}\exp\{\intop_{0}^{\ell}d\ell'\hat{H}(\ell')\}S_{p},
\end{equation}
but an explicit form of $\hat{H}(\ell$) is hardly accessible. However,
it is possible to encode $\hat{H}$ in terms of an infinite chain
of first-order equations for non-commutative Bell polynomials \citep{doi:10.1142/S0218196796000362}.
Indeed, given some invertible operator $\hat{{\cal U}}(\ell)$ that
is smooth in $\ell$ and $\hat{{\cal U}}(\ell=0)=1$, one can define
corresponding non-commutative Bell polynomials as solutions to
\begin{equation}
\frac{\partial}{\partial\ell}\hat{H}_{N}=\hat{H}_{N+1}-\hat{H}_{N}\hat{H}_{1},\quad\hat{H}_{N}(\ell=0)=(\frac{\partial}{\partial\ell})^{N}\hat{{\cal U}}_{\ell=0}.
\end{equation}
It is easy to check that explicit solutions are
\begin{equation}
\hat{H}_{N}=(\frac{\partial}{\partial\ell})^{N}\hat{{\cal U}}\hat{{\cal U}}^{-1}.
\end{equation}
Then $\hat{H}_{1}$ obeys
\begin{equation}
\frac{\partial}{\partial\ell}\hat{{\cal U}}=\hat{H}_{1}\hat{{\cal U}}
\end{equation}
and thus represents the Hamiltonian we look for. But to fully specify
the system, we need to find the Cauchy data, $(\frac{\partial}{\partial\ell})^{n}\hat{{\cal U}}_{\ell=0}$.

Let us represent $\Gamma_{p}$ as
\begin{equation}
\Gamma_{p}=\hat{{\cal U}}S_{p}=\sum_{n=0}^{\infty}\frac{(-i)^{n}}{n!}{\cal U}_{k_{1}...k_{n}}(\phi,\ell,\tau)\frac{\delta}{\delta\phi_{k_{1}}}...\frac{\delta}{\delta\phi_{k_{n}}}S_{p}.
\end{equation}
One can find an equation for ${\cal U}_{k_{1}...k_{n}}$ as follows.
Using the Fourier transform for $S_{p}$
\begin{equation}
S_{p}[\phi]=\int{\cal D}\xi\widetilde{S}_{p}[\xi]e^{i\xi_{q}\phi^{q}},
\end{equation}
 $\Gamma_{p}$ becomes
\begin{equation}
\Gamma_{p}=\hat{{\cal U}}S_{p}=\int{\cal D}\xi\widetilde{S}_{p}[\xi]e^{i\xi^{q}(\phi_{q}+\hbar\ell G_{qk}\frac{\delta}{\delta\phi_{k}})}1.
\end{equation}
Then, defining
\begin{equation}
F=\exp(-i\xi^{q}\phi_{q})\exp\{i\xi^{q}(\phi_{q}+\hbar\ell G_{qk}\frac{\delta}{\delta\phi_{k}})\}1,
\end{equation}
one has
\begin{equation}
\Gamma_{p}=\hat{{\cal U}}S_{p}=\int{\cal D}\xi\widetilde{S}_{p}[\xi]e^{i\xi_{q}\phi^{q}}F(\xi).
\end{equation}
and
\begin{equation}
\frac{\partial}{\partial\xi^{k}}F=i\hbar\ell G_{kq}\delta^{q}F-\hbar\ell G_{kq}\xi^{q}F,\quad F(\ell=0)=1.\label{F_eq}
\end{equation}
When deriving \eqref{F_eq}, one uses
\begin{equation}
[\phi_{q}+\hbar\ell G_{qk}\frac{\delta}{\delta\phi_{k}},\phi_{q'}+\hbar\ell G_{q'k'}\frac{\delta}{\delta\phi_{k'}}]=0
\end{equation}
which relies on the fact that $G_{qk}$ is the inverse to the second
variational derivative of the effective action.

Then, representing
\begin{equation}
F=\sum_{n=0}^{\infty}\frac{1}{n!}{\cal U}^{k_{1}...k_{n}}\xi_{k_{1}}...\xi_{k_{n}},\quad{\cal U}^{k_{1}...k_{n}}(\ell=0)=\delta_{n,0},\quad{\cal U}=1,
\end{equation}
and substituting this into \eqref{F_eq}, one finds, making use of
\begin{equation}
(\Gamma^{pq}+R^{pq})G_{qk}=\delta^{p}{}_{k},
\end{equation}
that
\begin{equation}
\hbar\ell\delta_{p}{\cal U}^{k_{1}...k_{n}}+i(2\pi)^{D}\hbar\ell n\delta_{p}{}^{k_{1}}{\cal U}^{k_{2}...k_{n}}+i(2\pi)^{D}(\Gamma_{pq}+R_{pq}){\cal U}^{qk_{1}...k_{n}}=0.\label{Cauchy_rel}
\end{equation}
Now we define
\begin{equation}
u_{p_{1}...p_{n}}^{(N)}=(\frac{\partial}{\partial\ell})^{N}{\cal U}_{p_{1}...p_{n}}(\ell=0)
\end{equation}
and from \eqref{Cauchy_rel} one obtains
\begin{equation}
\hbar\delta_{q}u_{p_{1}...p_{n}}^{(N)}+i\hbar n(2\pi)^{D}\delta_{qp_{1}}u_{p_{2}...p_{n}}^{(N)}+\frac{i(2\pi)^{D}}{(N+1)}\sum_{M=0}^{N}\left(\begin{array}{c}
N+1\\
M
\end{array}\right)u_{kp_{1}...p_{n}}^{(N+1-M)}(\frac{\partial}{\partial\ell})^{M}(\Gamma+R)^{k}{}_{q}|_{\ell=0}=0.\label{u_unf_eq}
\end{equation}
with
\begin{equation}
u_{p_{1}...p_{n}}^{(0)}=\delta_{n,0},\quad u^{(N)}=\delta_{N,0}.
\end{equation}
For \eqref{u_unf_eq} one needs
\begin{align}
 & (\frac{\partial}{\partial\ell})^{M}(\Gamma+R)_{k_{1}k_{2}}|_{\ell=0}=\sum_{S=1}^{M}\frac{(-1)^{S}}{(2\pi)^{DS}}\underset{i_{1}+...+i_{S}=M}{\sum}\nonumber \\
 & \frac{M!}{i_{1}!...i_{S}!}(\Gamma+R)_{k_{1}a}(\frac{\partial}{\partial\ell})^{i_{1}}G^{ab}(\Gamma+R)_{bc}...(\Gamma+R)_{xy}(\frac{\partial}{\partial\ell})^{i_{S}}G^{yz}(\Gamma+R)_{zk_{2}}|_{\ell=0},
\end{align}
which stems from the formula for a derivative of an inverse matrix.

Thus, the initial conditions for $\hat{H}_{N}$ are
\begin{equation}
\hat{H}_{N}(\ell=0)=\sum_{m=0}^{\infty}\frac{(-i)^{m}}{m!}u_{k_{1}...k_{m}}^{(N)}(\phi,\tau)\delta^{k_{1}}...\delta^{k_{m}}.
\end{equation}

Now we define a full propagator as
\begin{equation}
\frac{\partial}{\partial t}K_{p_{1}p_{2}}=-(\Gamma_{p_{1}}{}^{q}+R_{p_{1}}{}^{q})K_{qp_{2}},\quad K_{p_{1}p_{2}}(t=0)=(2\pi)^{D}\delta_{p_{1}p_{2}},\label{unf_heat_kernel}
\end{equation}
\begin{equation}
\frac{\partial}{\partial t}{\cal G}_{p_{1}p_{2}}=-K_{p_{1}p_{2}},\quad{\cal G}_{p_{1}p_{2}}(t\rightarrow\infty)=0,\label{unf_propagator}
\end{equation}
where a field-independent regulator $R_{p_{1}p_{2}}(\tau)$ depends
on a scale parameter $\tau$ referred to as \emph{RG-time}. These
equations are manifestly solved as
\begin{equation}
K_{p_{1}p_{2}}=(2\pi)^{D}e^{-t(\Gamma^{(2)}+R)}{}_{p_{1}p_{2}},\quad{\cal G}_{p_{1}p_{2}}=(2\pi)^{D}\frac{e^{-t(\Gamma^{(2)}+R)}}{\Gamma^{(2)}+R}{}_{p_{1}p_{2}}
\end{equation}
and we define
\begin{equation}
G_{p_{1}p_{2}}={\cal G}_{p_{1}p_{2}}(t=0).
\end{equation}

Finally, we formulate the full list of independent first-order equations
that determines a scalar QFT. One starts with a classical action $S[\phi]$
and a field-independent regulator $R_{p_{1}p_{2}}(\tau)$, then 
\begin{equation}
\delta_{p}S=S_{p},\quad\frac{\partial}{\partial\tau}S=\frac{\partial}{\partial\ell}S=0,\label{list_S}
\end{equation}
\begin{equation}
\frac{\partial}{\partial\tau}R_{p_{1}p_{2}}=R_{p_{1}p_{2}}^{(1)},\quad\delta_{q}R_{p_{1}p_{2}}=\frac{\partial}{\partial\ell}R_{p_{1}p_{2}}=0.\label{list_reg}
\end{equation}
Quantization is performed by the Schrödinger-like equation
\begin{equation}
\frac{\partial}{\partial\ell}\Gamma_{p}=\hat{H}_{1}\Gamma_{p},\quad\Gamma_{p}(\ell=0)=S_{p}\label{list_Schroedinger}
\end{equation}
where a formal Hamiltonian $\hat{H}_{1}$ is determined by

\begin{align}
 & \frac{\partial}{\partial\ell}\hat{H}_{N}=\hat{H}_{N+1}-\hat{H}_{N}\hat{H}_{1},\quad\hat{H}_{N}(\ell=0)=\sum_{n=0}^{\infty}\frac{(-i)^{n}}{n!}u_{k_{1}...k_{n}}^{(N)}\delta^{k_{1}}...\delta^{k_{n}};\label{list_Bell}\\
 & \hbar\delta_{q}u_{p_{1}...p_{n}}^{(N)}+i\hbar n(2\pi)^{D}\delta_{qp_{1}}u_{p_{2}...p_{n}}^{(N)}+\frac{i(2\pi)^{D}}{(N+1)}\sum_{M=0}^{N}\left(\begin{array}{c}
N+1\\
M
\end{array}\right)u_{kp_{1}...p_{n}}^{(N+1-M)}\sum_{S=1}^{M}(-1)^{S}\underset{i_{1}+...+i_{S}=M}{\sum}\nonumber \\
 & \frac{M!}{i_{1}!...i_{S}!}(\Gamma+R)^{k}{}_{a}G^{(i_{1})ab}(\Gamma+R)_{bc}...(\Gamma+R)_{xy}G^{(i_{S})yz}(\Gamma+R)_{zq}|_{\ell=0}=0;\label{list_u}\\
 & u_{p_{1}...p_{n}}^{(0)}=\delta_{n,0},\quad u^{(N)}=\delta_{N,0}.
\end{align}

The scale dependence of vertex functions is determined by the variational
derivative of the Wetterich equation \eqref{Wetterich_eq}, which,
considering the field independence of $R_{p_{1}p_{2}}$, is
\begin{equation}
\frac{\partial}{\partial\tau}\Gamma_{p}=\frac{\hbar}{2}G^{k_{1}k_{2}}{}_{|p}R_{k_{1}k_{2}}^{(1)}.\label{list_Wetterich}
\end{equation}
To close the system, one must define a two-point vertex $\Gamma_{p_{1}p_{2}}$
and a propagator $G_{p_{1}p_{2}}={\cal G}_{p_{1}p_{2}}(t=0)$ together
with its derivatives, which is achieved by
\begin{equation}
\delta_{q}\Gamma_{p}=\Gamma_{qp}.\label{list_two-point}
\end{equation}
\begin{equation}
\frac{\partial}{\partial t}K_{p_{1}p_{2}}=-(\Gamma_{p_{1}}{}^{q}+R_{p_{1}}{}^{q})K_{qp_{2}},\quad K_{p_{1}p_{2}}(t=0)=(2\pi)^{D}\delta_{p_{1}p_{2}}\label{list_kernel}
\end{equation}
\begin{equation}
\frac{\partial}{\partial t}{\cal G}_{p_{1}p_{2}}=-K_{p_{1}p_{2}},\quad{\cal G}_{p_{1}p_{2}}(t\rightarrow\infty)=0.\label{list_Green}
\end{equation}
\begin{equation}
\delta_{k}G_{p_{1}p_{2}}=G_{p_{1}p_{2}|k}.\label{list_variational_G}
\end{equation}
\begin{equation}
\frac{\partial}{\partial\ell}G_{p_{1}p_{2}}=G_{p_{1}p_{2}}^{(1)},\quad\frac{\partial}{\partial\ell}G_{p_{1}p_{2}}^{(n)}=G_{p_{1}p_{2}}^{(n+1)}.\label{list_derivative_G}
\end{equation}
Thus, the minimal set of unfolded fields one needs is:
\begin{enumerate}
\item classical action $S[\phi]$ and its 1-point vertex $S_{p}[\phi]$;
\item regulator $R_{p_{1}p_{2}}[\tau]$;
\item 1-point quantum vertex $\Gamma_{p}[\phi,\ell,\tau]$ and 2-point quantum
vertex $\Gamma_{p_{1}p_{2}}[\phi,\ell,\tau]$;
\item infinite sequence of non-commutative Bell polynomials $\hat{H}_{N}[\phi,\ell,\tau]$;
\item initial conditions for $\hat{H}_{N}[\phi,\ell,\tau]$ denoted by $u_{p_{1}...p_{n}}^{(N)}[\phi,\tau]$;
\item auxiliary heat kernel \textbf{$K_{p_{1}p_{2}}[\phi,\tau,t]$};
\item propagator defined as $G_{p_{1}p_{2}}[\phi,\ell,\tau]={\cal G}_{p_{1}p_{2}}[\phi,\ell,\tau,t=0]$
together with its $\ell$-derivatives $G_{p_{1}p_{2}}^{(n)}$ and
a variational derivative ${\cal G}_{p_{1}p_{2}|k}$.
\end{enumerate}
This completes the list of all fields and equations defining an unfolded
formulation of the Schwinger\textendash Dyson hierarchy. However,
to fully unfold the system, one must define, for every unfolded field,
its variational, $\ell$- and $\tau$-derivatives in terms of other
unfolded fields. The missing equations arise either from introducing
new unfolded descendant fields, as in
\begin{equation}
\delta_{p}\hat{H}_{N}=\hat{H}_{N|p},\quad\delta_{q}S_{p}=S_{qp},
\end{equation}
or from differentiating the equations already listed, as in
\begin{equation}
\frac{\partial}{\partial\ell}\Gamma_{p_{1}p_{2}}=\hat{H}_{1}\Gamma_{p_{1}p_{2}}+\hat{H}_{1|p_{1}}\Gamma_{p_{2}},\quad\Gamma_{p_{1}p_{2}}(\ell=0)=S_{p_{1}p_{2}},
\end{equation}
which follows from the variational differentiation of \eqref{list_Schroedinger}.

Since the descendant equations follow from the presented list of independent
equations, the resulting system is consistent provided that this list
is itself consistent. It is straightforward to see that this is indeed
the case. In fact, there is very little to check, since consistency
mostly imposes obvious constraints on unfolded descendant fields (e.g.,
the mutual consistency of the equations \eqref{list_reg} merely requires
that the $\tau$-descendants of $R_{p_{1}p_{2}}$ must also be $\phi$-
and $\ell$-independent). There is, however, one important nontrivial
constraint arising from the mutual consistency of \eqref{list_Schroedinger}
and \eqref{list_Wetterich}. It reads
\begin{equation}
(\frac{\partial}{\partial\tau}\hat{H}_{1})\Gamma_{p}-\frac{\hbar}{2}(\frac{\partial}{\partial\ell}G^{k_{1}k_{2}}{}_{|p})R_{k_{1}k_{2}}^{(1)}+\frac{\hbar}{2}(\hat{H}_{1}G^{k_{1}k_{2}}{}_{|p})R_{k_{1}k_{2}}^{(1)}=0\label{consistency_constraint}
\end{equation}
and must be taken into account when introducing descendant fields.
If the background is translation-invariant, then, in order to preserve
this symmetry, the regulator should take the form
\begin{equation}
R_{k_{1}k_{2}}=R(k,\tau)\delta_{k_{1}+k_{2}},\quad\Rightarrow\quad R_{k_{1}k_{2}}^{(1)}=\frac{\partial R(k,\tau)}{\partial\tau}\delta_{k_{1}+k_{2}}.
\end{equation}
Then, \eqref{consistency_constraint} can be solved as
\begin{equation}
\frac{\partial}{\partial\ell}G^{k_{1}k_{2}}{}_{|p}=\hat{H}_{1}G^{k_{1}k_{2}}{}_{|p}+\frac{2}{\hbar}\delta^{k_{1}k_{2}}V^{-1}(\frac{\partial R}{\partial\tau})^{-1}(\frac{\partial}{\partial\tau}\hat{H}_{1})\Gamma_{p}+F^{k_{1}k_{2}}{}_{|p},\label{solution_consistency}
\end{equation}
where $F^{k_{1}k_{2}}{}_{|p}$ is an unfolded descendant field subject
to
\begin{equation}
R_{k_{1}k_{2}}^{(1)}F^{k_{1}k_{2}}{}_{|p}=0,
\end{equation}
and $V$ is the regularized spacetime volume, defined by
\begin{equation}
\delta_{k_{1}+k_{2}}\delta^{k_{1}+k_{2}}=V,
\end{equation}
as is customary in FRG calculations.

Unfolding requires all equations to be formulated in terms of exterior
forms. For the problem in question, it means that we need to consider
1-forms on a base manifold which is ${\cal M}={\cal C}\times\mathbb{R}_{+}\times\mathbb{R}_{+}\times\mathbb{R}_{+}$
(three $\mathbb{R}_{+}$ factors correspond to $\tau$, $\ell$ and
$t$ coordinates). According to this, we introduce a full basis of
flat background vielbein 1-forms $e^{\hat{p}}$, $e^{(\tau)}$, $e^{(\ell)}$
and $e^{(t)}$ together with a connection 1-form $\omega^{\hat{p},\hat{q}}$,
where $\hat{p}$ stands for a tangent space momentum index.

The Cartan structure equations describe the flat background geometry
\begin{equation}
\mathrm{d}e{}^{\hat{p}}+\omega^{\hat{p},\hat{q}}e_{\hat{q}}=0,\quad\mathrm{d}\omega^{\hat{p},\hat{q}}+\omega^{\hat{p},}{}_{\hat{r}}\omega^{\hat{r},\hat{q}}=0,\quad\mathrm{d}e{}^{(\tau)}=0,\quad\mathrm{d}e{}^{(\ell)}=0,\quad\mathrm{d}e{}^{(t)}=0,\label{unf_background}
\end{equation}
where the exterior derivative on ${\cal M}$ is
\begin{equation}
\mathrm{d}=\delta\phi_{p}\frac{\delta}{\delta\phi_{p}}+\mathrm{d}\ell\frac{\partial}{\partial\ell}+\mathrm{d}\tau\frac{\partial}{\partial\tau}+\mathrm{d}t\frac{\partial}{\partial t}.
\end{equation}
After having determined $\phi$-, $\ell$- and $\tau$-derivatives
(also $t$-derivatives for $K$ and ${\cal G}$) for all unfolded
fields, one can write down a complete set of unfolded equations. For
any unfolded field $F$ with 
\begin{equation}
\delta_{p}F=F_{p},\quad\frac{\partial}{\partial\ell}F=F^{\ell},\quad\frac{\partial}{\partial\tau}F=F^{\tau}\quad\frac{\partial}{\partial t}F=F^{t},
\end{equation}
an unfolded equation is
\begin{equation}
\mathrm{d}_{\omega}F=e^{\hat{p}}F_{\hat{p}}+e^{(\ell)}F^{\ell}+e{}^{(\tau)}F^{\tau}+e{}^{(t)}F^{t}.\label{unf_eq_F}
\end{equation}
where a covariant derivative is defined as $\mathrm{d}_{\omega}=\mathrm{d}+\omega$.

Like any unfolded system, \eqref{unf_background}, \eqref{unf_eq_F}
possesses invariance with respect to background diffeomorphisms and
unfolded gauge symmetries \eqref{unf_gauge_transf}. In this case,
background diffeomorphism invariance means that the unfolded QFT system
is manifestly invariant under field redefinitions and coordinate changes
on $\ell$-, $\tau$- and $t$-rays. On the other hand, the unfolded
gauge symmetry here corresponds to local redefinitions that preserve
the flat fiber metric, in a straightforward analogy with the unfolded
scalar field of Section \ref{SECTION_UNFOLDING}. Thus, the final
unfolded formulation in terms of exterior forms is invariant under
an infinite-dimensional Euclidean algebra realized as linear fiber
field redefinitions (hatted indices) and under diffeomorphisms realized
as general nonlinear redefinitions of implicit base-manifold fields
(analogous to the coordinates of a background manifold in the scalar
field example of Section \ref{SECTION_UNFOLDING}). Fixing a particular
solution to the structure equations \eqref{unf_background} implements
the global choice of field variables and breaks the diffeomorphism
symmetry down to the global infinite-dimensional Euclidean redefinitions
realized in terms of these fields.

The initial equations are recovered from the unfolded system by choosing
Cartesian coordinates as a solution to the structure equations \eqref{unf_background}:
\begin{equation}
e_{0}^{\hat{p}}=\delta\phi^{\hat{p}},\quad\omega_{0}^{\hat{p},\hat{q}}=0,\quad e_{0}{}^{(\tau)}=\mathrm{d}\tau,\quad e_{0}{}^{(\ell)}=\mathrm{d}\ell,\quad e_{0}{}^{(t)}=\mathrm{d}t.
\end{equation}

\section{Example: $\lambda\phi^{4}$ theory in the one-loop approximation\label{SECTION_ONE_LOOP}}

To illustrate the structure of the unfolded QFT we constructed, let
us consider the scalar $\lambda\phi^{4}$ theory in $\mathbb{R}^{4}$
in the one-loop approximation. We start with the classical momentum-space
action
\begin{equation}
S[\phi]=\int\frac{\mathrm{d^{4}}p}{(2\pi)^{4}}\frac{1}{2}(p^{2}+m^{2})\phi(p)\phi(-p)+\frac{\lambda}{4!}\prod_{i=1}^{4}\int\frac{\mathrm{d^{4}}p_{i}}{(2\pi)^{4}}\phi(p_{i})(2\pi)^{4}\delta(p_{1}+p_{2}+p_{3}+p_{4}),
\end{equation}
which in our notation is
\begin{equation}
S[\phi]=\frac{1}{2}(p^{2}+m^{2})\phi^{p}\phi_{-p}+\frac{\lambda}{4!}\phi^{p_{1}}\phi^{p_{2}}\phi^{p_{3}}\phi^{p_{4}}(2\pi)^{4}\delta_{p_{1}+p_{2}+p_{3}+p_{4}}.\label{class_S}
\end{equation}
Now we analyze the list of independent unfolded equations \eqref{list_S}-\eqref{list_derivative_G}
up to $O(\hbar)$ order.

From \eqref{list_S} we find
\begin{equation}
S_{p}=(p^{2}+m^{2})\phi_{-p}+\frac{\lambda}{3!}\phi^{p_{1}}\phi^{p_{2}}\phi^{p_{3}}(2\pi)^{4}\delta_{p+p_{1}+p_{2}+p_{3}},
\end{equation}
and, for later use,
\begin{equation}
S_{p_{1}p_{2}}=\delta_{p_{1}}S_{p_{2}}=(2\pi)^{4}(p_{1}^{2}+m^{2})\delta_{p_{1}+p_{2}}+\frac{\lambda}{2}\phi^{p_{3}}\phi^{p_{4}}(2\pi)^{4}\delta_{p_{1}+p_{2}+p_{3}+p_{4}},
\end{equation}
\begin{equation}
S_{p_{1}p_{2}p_{3}}=\delta_{p_{1}}S_{p_{2}p_{3}}=\lambda\phi_{-p_{1}-p_{2}-p_{3}},
\end{equation}
\begin{equation}
S_{p_{1}p_{2}p_{3}p_{4}}=\delta_{p_{1}}S_{p_{2}p_{3}p_{4}}=\lambda(2\pi)^{4}\delta_{p_{1}+p_{2}+p_{3}+p_{4}}.
\end{equation}
We keep the regulator $R_{p_{1}p_{2}}$ arbitrary for now and come
back to \eqref{list_reg} later. First, let us analyze the system
in the classical $\hbar=0$ limit. To solve \eqref{list_Schroedinger},
we need to find the Hamiltonian $\hat{H}_{1}$ from \eqref{list_Bell},
which requires solving \eqref{list_u}. At $\hbar=0$, from \eqref{list_u}
one finds
\begin{equation}
u_{p_{1}...p_{n}}^{(N)}(\hbar=0)=\delta_{N,0}\delta_{n,0}.
\end{equation}
Then \eqref{list_Bell} yields
\begin{equation}
\hat{H}_{N}(\hbar=0)=\delta_{N,0}
\end{equation}
and from \eqref{list_Schroedinger}
\begin{equation}
\Gamma_{p}(\hbar=0)=S_{p}.
\end{equation}
Now \eqref{list_kernel}, \eqref{list_Green} give
\begin{equation}
K_{p_{1}p_{2}}(\hbar=0)=\exp\{-t(S^{(2)}+R)\}_{p_{1}p_{2}},\quad{\cal G}_{p_{1}p_{2}}(\hbar=0)=(2\pi)^{4}\frac{\exp\{-t(S^{(2)}+R)\}}{(S^{(2)}+R)}_{p_{1}p_{2}}.
\end{equation}
This is all the classical information we need.

At the one-loop order, \eqref{list_u} is solved by
\begin{equation}
u_{p_{1}...p_{n}}^{(N)}=\delta_{N,0}\delta_{n,0}-\hbar\delta_{n,2}\delta_{N,1}G_{p_{1}p_{2}}(\hbar=0)+O(\hbar^{2}),
\end{equation}
which leads via \eqref{list_Bell} to
\begin{equation}
\hat{H}_{N}=\delta_{N,0}+\frac{\hbar}{2}\delta_{N,1}G^{p_{1}p_{2}}(\hbar=0)\delta_{p_{1}}\delta_{p_{2}}+O(\hbar^{2}).
\end{equation}
Then the solution to \eqref{list_Schroedinger} is
\begin{equation}
\Gamma_{p}=S_{p}+\frac{\hbar\ell}{2}G^{p_{1}p_{2}}(\hbar=0)S_{pp_{1}p_{2}}+O(\hbar^{2}).\label{Gamma_h}
\end{equation}
Taking variational derivatives of \eqref{Gamma_h}, one finds
\begin{equation}
\Gamma_{p_{1}p_{2}}=S_{p_{1}p_{2}}+\frac{\hbar\ell}{2}G^{k_{1}k_{2}}(\hbar=0)S_{p_{1}p_{2}k_{1}k_{2}}+\frac{\hbar\ell}{2}G^{k_{1}k_{2}}{}_{|p_{1}}(\hbar=0)S_{p_{2}k_{1}k_{2}}+O(\hbar^{2}).\label{Gamma_2_h}
\end{equation}

\begin{equation}
\Gamma_{p_{1}p_{2}p_{3}p_{4}}=S_{p_{1}p_{2}p_{3}p_{4}}+\frac{3\hbar\ell}{2}G^{k_{1}k_{2}}{}_{|p_{1}p_{2}}(\hbar=0)S_{k_{1}k_{2}p_{3}p_{4}}+O(\hbar^{2}).\label{Gamma_4_h}
\end{equation}
Without regularization, both expressions \eqref{Gamma_2_h} and \eqref{Gamma_4_h}
diverge. Now we make use of the freedom in choosing $R_{p_{1}p_{2}}$
and take 
\begin{equation}
R_{p_{1}p_{2}}(\tau)=(2\pi)^{4}(e^{\tau(p_{1}^{2}+m^{2})}-1)(p_{1}^{2}+m^{2})\delta_{p_{1}+p_{2}}.
\end{equation}
This introduces heat-kernel regularization and makes all loop integrals
UV-finite, yielding
\begin{align}
\Gamma_{p_{1}p_{2}}[\phi=0]= & (2\pi)^{4}(p_{1}^{2}+m^{2})\delta_{p_{1}+p_{2}}+\frac{\hbar\ell\lambda m^{2}}{32\pi^{2}}\Gamma(-1,\tau m^{2})\delta_{p_{1}+p_{2}}+O(\hbar^{2}),\label{Gamma_2}\\
\Gamma_{p_{1}p_{2}p_{3}p_{4}}[\phi=0]= & \lambda(2\pi)^{4}\delta_{\sum p}+\frac{\hbar\ell(-\lambda^{2})}{16\pi^{2}}\delta_{\sum p}\int_{0}^{1}\frac{dx}{(1+x)^{2}}\Gamma\Bigl(0,[\frac{m^{2}(1+x)}{x}+\frac{s}{1+x}]\tau\Bigr)+\nonumber \\
 & +(t\text{-channel})+(u\text{-channel})+O(\hbar^{2}).\label{Gamma_4}
\end{align}

To provide a physical interpretation of the obtained unfolded vertex
functions, one needs to give a concrete meaning to the RG-time $\tau$.
Leaving a systematic study of the problem of renormalization of unfolded
QFT to future work, here we restrict ourselves to stating that, with
the identification $\tau=1/\Lambda^{2}$ for a UV cutoff $\Lambda$,
the standard one-loop beta function can be recovered as
\begin{equation}
\beta(\lambda)=\lim_{\tau\rightarrow0}2\tau\frac{\partial}{\partial\tau}\Gamma^{(4)}=\frac{3\hbar\ell\lambda^{2}}{16\pi^{2}}+O(\lambda^{3}).
\end{equation}

\section{Conclusion\label{SECTION_CONCLUSION}}

In this paper, we have constructed an unfolded formulation of the
Schwinger\textendash Dyson hierarchy governing the tower of vertex
functions of a scalar QFT. Our construction invokes two additional
evolution parameters that govern WKB and RG flows. As an illustration,
we performed a one-loop analysis of $\lambda\phi^{4}$ theory.

The system we built poses a lot of questions for future studies. The
most straightforward problem is to relate the proposed construction,
formulated in terms of vertex functions, to unfolded\emph{ field }equations
(for the unfolded variational problem see \citep{Tarusov:2021fdk}).
A related question addresses the UV regulator we made use of in the
$\lambda\phi^{4}$ example: an analogous heat-kernel contribution
was introduced in \citep{Misuna:2019ijn} (see also \citep{Misuna:2020fck,Misuna:2022zjr})
in order to provide an off-shell completion of classical on-shell
unfolded field equations. Then in \citep{Misuna:2022cma} it was demonstrated
that the evolution parameter of this heat kernel plays the crucial
role for quantization of unfolded field equations, determining the
unfolded field-source commutator, and serves as a built-in UV regulator,
smearing the delta function that arises in this commutator. In addition,
a similar heat-kernel convolution was introduced in \citep{Luscher:2010iy}
(see also \citep{Luscher:2011bx,Suzuki:2013gza,Carosso:2018rep,Makino:2018rys,Abe:2018zdc})
to define \emph{the gradient flow} in lattice QCD. It is interesting
to find the relation between these constructions.

Other prominent directions include development of renormalization
techniques (in particular, specifying the connection to functional
RG \citep{Wetterich:1989xg,Wetterich:1992yh,Tetradis:1993ts,Pawlowski:2005xe,Dupuis:2020fhh}),
generalization to gauge theories and introducing nontrivial geometry
on the configuration space (especially interesting in the context
of RG analysis of nonlinear sigma-models \citep{Friedan:1980jf,Flore:2012ma,Efremov:2021fub,Haruna:2024lmj}).

One more interesting direction is the unfolded analysis of the $\tau$-
and $\ell$-flows in the spirit of the Q-cohomology technique of \citep{Vasiliev:2005zu}.
Corresponding $\tau$-cohomologies then would correspond to RG invariants,
while $\ell$-cohomologies would describe non-renormalized operators.

\section*{Acknowledgments}

The author is grateful to A.V. Korybut, A.S. Koshelev, N.V. Krasnikov,
A.V. Marshakov, E.T. Musaev and M.A. Vasiliev for valuable comments
and remarks. During the preparation of this manuscript, the author
used ChatGPT (OpenAI, GPT-6) to identify relevant literature, improve
the English, assist with the analytical evaluation of integrals in
Section 5, and provide critical feedback on successive drafts. The
author takes full responsibility for the scientific content, results,
and conclusions of the manuscript. The work was supported by RSF grant
26-12-00240.

\printbibliography

\end{document}